\documentclass[11pt,english]{article}
\usepackage{amssymb, amsmath}
\usepackage{multicol}
\usepackage{amsmath, amssymb}
\usepackage[mathscr]{eucal}
\usepackage{graphicx}
\usepackage{amsfonts}
\usepackage{babel}
\newcommand{\bfi}{\bfseries\itshape}
\newcommand{\rem}[1]{}
\newcommand{\remfigure}[1]{}
\makeatletter
\@addtoreset{equation}{section}

\makeatother

\newtheorem{theorem}{Theorem}

\numberwithin{theorem}{section}

\def\0{{\bf 0}}

\begin{document}

\title{
Singular solutions for geodesic flows of Vlasov moments}
\author{\vspace{2mm}
J. Gibbons$^{1}$, D. D. Holm$^{1,\,2}$ and C. Tronci$^{1,\,3}$\\
$^1$ Department of Mathematics, Imperial College London, London SW7 2AZ, UK\,.\\
$^2$ Computer and Computational Science Division, Los Alamos National Laboratory, \\Los Alamos, NM, 87545 USA 
\\$^3$\,TERA Foundation for Oncological Hadrontherapy\\
11 Via Puccini, Novara 28100, Italy\,\\ \\
}
\date{For Henry McKean, on the occasion of his 75th birthday\\
}

\maketitle

\begin{abstract}\noindent 
The Vlasov equation for the collisionless evolution of the single-particle
probability distribution function (PDF) is a well-known example of
coadjoint motion. Remarkably, the property of coadjoint motion survives
the process of taking moments. That is, the evolution of the moments of
the Vlasov PDF is also coadjoint motion. We find that {\it geodesic}
coadjoint motion of the Vlasov moments with respect to powers of the
single-particle momentum admits singular (weak) solutions concentrated on
embedded subspaces of physical space. The motion and interactions of these
embedded subspaces are governed by canonical Hamiltonian equations for
their geodesic evolution. 
\rem{
By symmetry under exchange of canonical
momentum and position, the Vlasov moments with respect to powers of the
single-particle position admit singular (weak) solutions concentrated on
embedded subspaces of momentum space.}
\end{abstract}

\tableofcontents

\section{Introduction}

\paragraph{The Vlasov equation.}

The evolution of $N$ identical particles in phase space with coordinates 
$(q_i, p_i)$ $i=1,2,\dots,N$, may be described by an evolution equation
for their joint probability distribution function. Integrating over all
but one of the particle phase-space coordinates yields an evolution
equation for the single-particle probability distribution function (PDF).
This is the Vlasov equation. 

The solutions of the Vlasov equation reflect its heritage in particle
dynamics, which may be reclaimed by writing its many-particle PDF as a
product of delta functions in phase space. Any number of these delta
functions may be integrated out until all that remains is the dynamics of
a single particle in the collective field of the others. In plasma
physics, this collective field generates the total electromagnetic
properties and the self-consistent equations obeyed by the single
particle PDF are the Vlasov-Maxwell equations. In the electrostatic
approximation, these become the Vlasov-Poisson equations, which govern
the statistical distributions of particle systems ranging from integrated
circuits (MOSFETS, metal-oxide semiconductor field-effect transistors), to
charged-particle beams, to the distribution of galaxies in the Universe. 

A class of singular solutions of the VP equations called the ``cold
plasma'' solutions have a particularly beautiful experimental realization
in the Malmberg-Penning trap. In this experiment, the time average of
the vertical motion closely parallels the Euler fluid equations. In
fact, the cold plasma singular Vlasov-Poisson solution turns out to obey
the equations of point-vortex dynamics in an incompressible ideal flow.
This coincidence allows the discrete arrays of ``vortex crystals''
envisioned by J. J. Thomson for fluid vortices to be realized
experimentally as solutions of the Vlasov-Poisson equations. For a
survey of these experimental cold-plasma results see \cite{DuON1999}. 

\paragraph{Vlasov moments.}

The Euler fluid equations arise by imposing a closure relation
on the first three momentum moments, or $p-$moments of the Vlasov PDF
$f(p,q,t)$. The zero-th $p-$moment is the spatial density of particles.
The first $p-$moment is the mean momentum and its ratio with the zero-th
$p-$moment is the Eulerian fluid velocity.  Introducing an expression for the fluid pressure in terms of the density and velocity closes the system of $p-$moment equations, which otherwise would possess a countably infinite number of dependent variables. 

The operation of taking $p-$moments preserves the geometric nature of
Vlasov's equation. It's closure after the first $p-$moment results in
Euler's useful and beautiful theory of ideal fluids. As its primary geometric
characteristic, Euler's fluid theory represents fluid flow as
Hamiltonian geodesic motion on the space of smooth invertible maps
acting on the flow domain and possessing smooth inverses. These smooth
maps (called diffeomorphisms) act on the fluid reference configuration so
as to move the fluid particles around in their container. And their smooth
inverses recall the initial reference configuration (or label) for the
fluid particle currently occupying any given position in space. Thus, the
motion of all the fluid particles in a container is represented as a
time-dependent curve in the infinite-dimensional group of
diffeomorphisms. Moreover, this curve describing the sequential actions
of the diffeomorphisms on the fluid domain is a special optimal curve that
distills the fluid motion into a single statement. Namely, ``A fluid moves
to get out of its own way as efficiently as possible.'' Put more
mathematically, fluid flow occurs along a curve in the diffeomorphism
group which is a geodesic with respect to the metric on its tangent space
supplied by its kinetic energy. 

Given the beauty and utility of the solution behavior for Euler's
equation for the first $p-$moment, one is intrigued to know more about
the dynamics of the other moments of Vlasov's equation. Of course, the
dynamics of the the $p-$moments of the Vlasov-Poisson equation is one of  the
mainstream subjects of plasma physics and space physics. 

\paragraph{Summary.}
This paper formulates the problem of Vlasov $p-$moments governed by {\it quadratic}
Hamiltonians. This dynamics is a certain type of geodesic motion on the
symplectomorphisms, rather than the diffeomorphisms. The
symplectomorphisms are smooth invertible maps acting on the phase
space and possessing smooth inverses. We shall consider the singular
solutions of the geodesic dynamics of the Vlasov $p-$moments. Remarkably,
these equations turn out to be related to integrable systems governing shallow water
wave theory. In fact, when the Vlasov $p-$moment equations for geodesic
motion on the symplectomorphisms are closed at the level of the first
$p-$moment, their singular solutions are found to recover the peaked soliton of the
integrable Camassa-Holm equation for shallow water waves \cite{CaHo1993}. 

Thus, geodesic symplectic dynamics of the Vlasov $p-$moments is found to
possess singular solutions whose closure at the fluid level recovers the
peakon solutions of shallow water theory. Being solitons, the peakons
superpose and undergo elastic collisions in fully nonlinear interactions.
The singular solutions for Vlasov $p-$moments presented here also
superpose and interact nonlinearly as coherent structures. \\
\rem{
However, they
differ qualitatively from the singular solutions of both the Camassa-Holm
equation and the Vlasov equation. This difference arises because the
momentum of the $j-$th singular solution for the $p-$moments is a
function of its corresponding canonically conjugate single-particle
momentum, as well as being a function of time.} 

\noindent
The plan of the paper follows: 
\begin{description}
\item
Section \ref{pmom-eqns} defines the Vlasov $p-$moment equations and
formulates them as Hamiltonian system using the Kupershmidt-Manin
Lie-Poisson bracket. This formulation identifies the $p-$moment
equations as coadjoint motion on a dual Lie algebra
$\mathfrak{g}^*$, in any number of spatial dimensions.
\item
Section \ref{VarPrincHP} derives variational formulations of the
$p-$moment dynamics in both their Lagrangian and Hamiltonian forms. 
\item
Section \ref{geo-prob} formulates the problem of geodesic motion on the
symplectomorphisms in terms of the Vlasov $p-$moments and identifies the
singular solutions of this problem, whose support is concentrated on
delta functions in position space. In a special case, the truncation of
geodesic symplectic motion to geodesic diffeomorphic motion for
the first $p-$moment recovers the singular solutions of the Camassa-Holm
equation. 
\item
Section \ref{geo-singsoln} discusses how the singular $p-$moment solutions
for geodesic symplectic motion are related to the cold plasma solutions.
By symmetry under exchange of canonical momentum $p$ and position $q$, the
Vlasov $q-$moments are also found to admit singular (weak) solutions.
\end{description}

\section{Vlasov moment dynamics}\label{pmom-eqns}

The Vlasov equation may be expressed as 
\begin{eqnarray}
\frac{\partial f}{\partial t}
=
\Big[f\,,\,\frac{\delta h}{\delta f}\Big]
=
\frac{\partial f}{\partial p}\frac{\partial}{\partial q}
\frac{\delta h}{\delta f}
-
\frac{\partial f}{\partial q}\frac{\partial}{\partial p}
\frac{\delta h}{\delta f}
=: -\,{\rm ad}^*_{\delta h/\delta f}\,f
\label{vlasov-eqn}
\end{eqnarray}
Here the canonical Poisson bracket $[\,\cdot\,,\,\cdot\,]$
is defined for smooth functions on phase space with
coordinates $(q,p)$ and $f(q,p,t)$ is the evolving Vlasov
single-particle distribution function.  The variational
derivative $\delta h/\delta f$ is the single
particle Hamiltonian. 

A functional $g[f]$ of the Vlasov distribution $f$ evolves
according to 
\begin{eqnarray*}
\frac{dg}{dt}
&=&
\int\hspace{-3mm}\int
\frac{\delta g}{\delta f}\,
 \frac{\partial f}{\partial t} 
\,dqdp 
=
\int\hspace{-3mm}\int
\frac{\delta g}{\delta f}\,
\Big[f\,,\,\frac{\delta h}{\delta f}\Big]
\,dqdp 
\\
&=&
-\int\hspace{-3mm}\int f 
\Big[\frac{\delta g}{\delta f}\,,\,\frac{\delta h}{\delta f}\Big]
\,dqdp 
=:-\,
\Big\langle\!\!\Big\langle f\,,\,
\Big[\frac{\delta g}{\delta f}\,,\,\frac{\delta h}{\delta f}\Big]
\Big\rangle\!\!\Big\rangle
=:
\{\,g\,,\,h\,\}
\end{eqnarray*}
In this calculation boundary terms are neglected upon integrating by
parts and the notation
$\langle\!\langle\,\cdot\,,\,\cdot\,\rangle\!\rangle$ is introduced for
the $L^2$ pairing in phase space. The quantity $\{\,g\,,\,h\,\}$ defined
in terms of this pairing is the Lie-Poisson Vlasov (LPV) bracket
\cite{MoWe}. This Hamiltonian evolution equation may also be expressed as
\begin{eqnarray*}
\frac{dg}{dt}
=
\{\,g\,,\,h\,\}
=
\Big\langle\!\!\Big\langle f\,,\,
{\rm ad}\,_{\delta h/\delta f}
\frac{\delta g}{\delta f}
\Big\rangle\!\!\Big\rangle
=
-\,
\Big\langle\!\!\Big\langle 
{\rm ad}^*\,_{\delta h/\delta f}\,f\,,\,
\frac{\delta g}{\delta f}
\Big\rangle\!\!\Big\rangle
\end{eqnarray*}
which defines the Lie-algebraic operations ad and ad$^*$ in this
case in terms of the $L^2$ pairing on phase space
$\langle\!\langle\,\cdot\,,\,\cdot\,\rangle\!\rangle$:
$\mathfrak{s}^*\times\mathfrak{s}\mapsto\mathbb{R}$. Thus, the
notation ${\rm ad}^*_{\delta h/\delta f}\,f$ in
(\ref{vlasov-eqn}) expresses {\bfi coadjoint action} of $\delta
h/\delta f\in\mathfrak{s}$ on $f\in\mathfrak{s}^*$, where
$\mathfrak{s}$ is the Lie algebra of single particle Hamiltonian
vector fields and $\mathfrak{s}^*$ is its dual under $L^2$
pairing in phase space. This is the sense in which the Vlasov equation
represents coadjoint motion on the symplectomorphisms.

\subsection{Dynamics of Vlasov $q,p-$Moments}
The phase space $q,p-$moments of the Vlasov distribution function are
defined by 
\begin{eqnarray*}
g_{\,\widehat{m}m}
&=&
\int\hspace{-3mm}\int f (q,p)\,
q^{\widehat{m}}p^m
\,dq\,dp 
\,.
\end{eqnarray*}
The $q,p-$moments $g_{\,\widehat{m}m}$ are often used in
treating the collisionless dynamics of plasmas and particle beams
\cite{Dragt}. This is usually done by considering low order
truncations of the potentially infinite sum over phase space
moments,  
\begin{eqnarray*}
g
&=&
\sum_{\widehat{m},m=0}^\infty
a_{\,\widehat{m}m}g_{\,\widehat{m},m}
\,,\qquad
h
=
\sum_{\widehat{n},n=0}^\infty
b_{\,\widehat{n}n}g_{\,\widehat{n},n}
\,,
\end{eqnarray*}
with constants $a_{\,\widehat{m}m}$ and $b_{\,\widehat{n}n}$,
with $\widehat{m},m,\widehat{n},n=0,1,\dots$. If $h$ is the
Hamiltonian, the sum over $q,p-$moments $g$ evolves under the Vlasov
dynamics according to the Poisson bracket relation
\begin{eqnarray*}
\frac{dg}{dt}
=
\{\,g\,,\,h\,\}
&=&
\sum_{\widehat{m},m,\widehat{n},n=0}^\infty
a_{\,\widehat{m}m}b_{\,\widehat{n}n}
(\widehat{m}m-\widehat{n}n)
g_{\,\widehat{m}+\widehat{n}-1,m+n-1}
\,.
\end{eqnarray*}
The symplectic invariants associated with Hamiltonian
flows of the $q,p-$moments were discovered and classified
in \cite{HoLySc1990}.

\subsection{Dynamics of Vlasov $p-$Moments} 

The momentum moments, or ``$p-$moments,'' of the Vlasov function are
defined as
\begin{eqnarray*}
A_m(q,t)=\int p^m\,f(q,p,t)\,dp
\,,\qquad
m=0,1,\dots.
\end{eqnarray*}
 That is, the $p-$moments are $q-$dependent integrals over $p$
 of the product of powers $p^m$, $m=0,1,\dots$, times the
 Vlasov solution $f(q,p,t)$.
We shall consider functionals of these $p-$moments defined by,
\begin{eqnarray*}
g
&=&
\sum_{m=0}^\infty
\int\hspace{-3mm}\int
 \alpha_m(q)\,p^m\,f\,dqdp
=
\sum_{m=0}^\infty
\int
 \alpha_m(q)\,A_m(q)\,dq
=:
\sum_{m=0}^\infty\Big\langle A_m\,,\,\alpha_m\Big\rangle
\\
h
&=&
\sum_{n=0}^\infty
\int\hspace{-3mm}\int
 \beta_n(q)\,p^n\,f\,dqdp
=
\sum_{n=0}^\infty
\int
 \beta_n(q)\,A_n(q)\,dq
=:
\sum_{n=0}^\infty\Big\langle A_n\,,\,\beta_n\Big\rangle
\end{eqnarray*}
where $\langle\,\cdot\,,\,\cdot\,\rangle$ is the
$L^2$ pairing on position space.

The functions $\alpha_m$
and $\beta_n$ with $m,n=0,1,\dots$ are assumed to be suitably
smooth and integrable against the Vlasov $p-$moments. To
assure these properties, one may relate the $p-$moments to the
previous sums of Vlasov $q,p-$moments by choosing
\begin{eqnarray*}
\alpha_m(q)
&=&
\sum_{\widehat{m}=0}^\infty
a_{\,\widehat{m}m}q^{\,\widehat{m}}
\,,\qquad
\beta_n(q)
=
\sum_{\widehat{n}=0}^\infty
b_{\,\widehat{n}n}q^{\,\widehat{n}}
\end{eqnarray*}
For these choices of $\alpha_m(q)$ and $\beta_n(q)$, the sums of 
$p-$moments will recover the full set of Vlasov $(q,p)-$moments.
Thus, as long as the $q,p-$moments of the distribution $f(q,p)$
continue to exist under the Vlasov evolution, one may assume that the
dual variables $\alpha_m(q)$ and $\beta_n(q)$ are smooth functions whose
Taylor series expands the $p-$moments in the $q,p-$moments. These functions
are dual to the $p-$moments $A_m(q)$ with $m=0,1,\dots$ under the $L^2$
pairing $\langle\cdot\,,\,\cdot\rangle$ in the spatial variable $q$.
In what follows we will assume {\it homogeneous} boundary conditions. This means, for example, that we will ignore boundary terms arising from integrations by parts.

\subsection{Poisson bracket for Vlasov $p-$moments}

The Poisson bracket among the
$p-$moments is obtained from the LPV bracket via the following
explicit calculation, 
\begin{eqnarray*}
\{\,g\,,\,h\,\}
&=&
-\sum_{m,n=0}^\infty
\int\hspace{-3mm}\int f 
\Big[\alpha_m(q)\,p^m\,,\,\beta_n(q)\,p^n\Big]
\,dqdp 
\\
&=&
-\sum_{m,n=0}^\infty
\int\hspace{-3mm}\int 
\Big[
m\alpha_{m}\beta_{n}\,^{\prime}\,-n\beta_{n}\alpha_{m}\,^{\prime}\Big]
f\,p^{m+n-1}
\,dqdp 
\\
&=&
-\sum_{m,n=0}^\infty
\int
A_{m+n-1}(q)
\Big[
m\alpha_{m}\beta_{n}\,^{\prime}\,-n\beta_{n}\alpha_{m}\,^{\prime}\Big]
\,dq
\\
&=:&
\sum_{m,n=0}^\infty\Big\langle 
A_{m+n-1}
\,,\,
{\rm ad}_{\beta_n}\alpha_m
\Big\rangle 
\\
&=&
-\sum_{m,n=0}^\infty
\int
\Big[
n\beta_n A_{m+n-1}'
+(m+n)A_{m+n-1}\beta_n\,'\Big]
\alpha_m
\,dq
\\
&=:&
-\sum_{m,n=0}^\infty\Big\langle 
{\rm ad}^*_{\beta_n}A_{m+n-1}
\,,\,
\alpha_m
\Big\rangle 
\end{eqnarray*}
where we have integrated by parts and introduced the 
notation ad and ad$^*$ for adjoint and coadjoint action,
respectively. Upon recalling the dual relations 
\begin{eqnarray*}
\alpha_m=\frac{\delta g}{\delta A_m}
\quad\hbox{and}\quad
\beta_n=\frac{\delta h}{\delta A_n}
\end{eqnarray*}
the LPV bracket in terms of the $p-$moments may be expressed as
\begin{eqnarray*}
\{\,g\,,\,h\,\}(\{A\})
&=&
-\sum_{m,n=0}^\infty
\int
\frac{\delta g}{\delta A_m}
\Big[
n\frac{\delta h}{\delta A_n} 
\frac{\partial}{\partial q} A_{m+n-1}
+
(m+n)A_{m+n-1}\frac{\partial}{\partial q}
\frac{\delta h}{\delta A_n}\Big]
\,dq
\\
&=:&
-\sum_{m,n=0}^\infty
\Big\langle 
A_{m+n-1}
\,,\,
\Big[\!\!\Big[\frac{\delta g}{\delta A_m}\,,\,
\frac{\delta h}{\delta A_n}\Big]\!\!\Big]
\Big\rangle 
\end{eqnarray*}
This is the Kupershmidt-Manin Lie-Poisson (KMLP) bracket
\cite{KuMa}, which is defined for
functions on the dual of the Lie algebra with bracket
\begin{eqnarray*}
[\![\,\alpha_m\,,\,\beta_n\,]\!]
=
m\alpha_m\partial_q\beta_n-n\beta_n\partial_q\alpha_m
\,.
\end{eqnarray*}
This Lie algebra bracket inherits the Jacobi identity
from its definition in terms of the canonical Hamiltonian
vector fields. Thus, we have shown the
\begin{theorem}[Gibbons \cite{Gi}]$\quad$\\
The operation of taking
$p-$moments of Vlasov solutions is a Poisson map. It takes the LPV bracket
describing the evolution of $f(q,p)$ into the KMLP bracket, describing the
evolution of the $p-$moments $A_n(x)$. A result related to this, for
the Benney hierarchy \cite{Be1973}, was also noted by Lebedev and Manin
\cite{LeMa}.
\end{theorem}

The evolution of a particular $p-$moment $A_m(q,t)$ is obtained
from the KMLP bracket by
\begin{eqnarray*}
\frac{\partial  A_m}{\partial t}
=
\{\,A_m\,,\,h\,\}
&=&
-\sum_{n=0}^\infty
\Big(
n\frac{\delta h}{\delta A_n} 
\frac{\partial}{\partial q} A_{m+n-1}
+
(m+n)A_{m+n-1}\frac{\partial}{\partial q}
\frac{\delta h}{\delta A_n}
\Big)
\end{eqnarray*}
The KMLP bracket among the $p-$moments is given by
\begin{eqnarray*}
\{\,A_m\,,\,A_n\,\}
&=&
-n\frac{\partial}{\partial q} A_{m+n-1}
-
mA_{m+n-1}\frac{\partial}{\partial q}
\end{eqnarray*}
expressed as a differential operator acting to the
right. This operation is skew-symmetric under the $L^2$  pairing and the
general KMLP bracket can then be written as \cite{Gi}
\[
\{\,g\,,h\,\}\left(  \,\left\{  A\right\}  \right)  =\sum_{m,n=0}^{\infty}
\int\frac{\delta g}{\delta A_{m}}\{\,A_{m}\,,\,A_{n}\,\}\frac{\delta h}{\delta
A_{n}}dq
\]
so that
\[
\frac{\partial A_{m}}{\partial t}=\sum_{n=0}^{\infty}\{\,A_{m}\,,\,A_{n}
\,\}\frac{\delta h}{\delta A_{n}}.
\]

\subsection{Multidimensional treatment}
We now show that the KMLP bracket and the equations of motion may be written in three dimensions in multi-index notation.
By writing $\mathbf{p}^{2n+1}=\,p^{2n}\,  \mathbf{p}$, and checking that:
\begin{align*}
p^{2n} 
& =
\underset{i+j+k=n}{\sum}
\dfrac{n!}{i!j!k!}\,p_{1}^{2i}p_{2}^{2j}p_{3}^{2k}\\
\end{align*}
it is easy to see that the multidimensional treatment can be performed in terms of the quantities
\[
p^{\sigma}
=:
p_{1}^{\sigma_{1}}p_{2}^{\sigma_{2}}p_{3}^{\sigma_{3}}
\]
\bigskip where $\sigma\in\mathbb{N}^{3}$. Let $A_{\sigma}$ be defined as
\[
A_{\sigma}
\left(  
\mathbf{q},t\right)  
=:
{\int}
p^{\sigma}
f\left(\mathbf{q},\mathbf{p},t\right)
dp
\]
and consider functionals of the form
\begin{align*}
g  
& =
{\sum\limits_{\sigma}}
{\iint}
\alpha_{\sigma}\left(  \mathbf{q}\right)  
p^{\sigma}
f\left(  \mathbf{q},\mathbf{p},t\right)
dqdp=:
{\sum\limits_{\sigma\in\mathbb{N}^{3}}}
\left\langle 
A_{\sigma},
\alpha_{\sigma}
\right\rangle \\
h  
& =
{\sum\limits_{\rho}}
{\iint}
\beta_{\rho}
\left(  \mathbf{q}\right)  
p^{\,\rho}
f\left(  \mathbf{q},\mathbf{p},t\right)
dqdp=:
{\sum\limits_{\rho\in\mathbb{N}^{3}}}
\left
\langle A_{\rho},
\beta_{\rho}
\right\rangle
\end{align*}
The ordinary LPV bracket leads to: 

\begin{align*}
\left\{ g,h\right\} 
& =
-
\sum\limits_{\sigma ,\rho }\iint 
f
\left[ 
\alpha _{\sigma }
\left( 
\mathbf{q}
\right) 
p^{\sigma },
\beta _{\rho}
\left( 
\mathbf{q}
\right)
p^{\,\rho }
\right] 
dqdp
=
\\
& =
-
\sum\limits_{\sigma ,\rho }\sum\limits_{j}
\iint
f
\left( 
\alpha _{\sigma }
p^{\,\rho }
\dfrac{\partial p^{\sigma }}{\partial p_{j}}
\dfrac{\partial \beta _{\rho }}{\partial q_{j}}
-
\beta _{\rho }
p^{\sigma }
\dfrac{\partial p^{\,\rho }}{\partial p_{j}}
\dfrac{\partial \alpha _{\sigma }}{\partial q_{j}}
\right) 
dqdp
= 
\\
& =
-
\sum\limits_{\sigma ,\rho }\sum\limits_{j}
\iint
f
\left( 
\sigma _{j}
\alpha _{\sigma }\,
p^{\,\rho }\,
p^{\sigma -1_{\hbox{\normalsize\it j}}}\,
\dfrac{\partial \beta _{\rho }}{\partial q_{j}}
-
\rho _{j}
\beta _{\rho }\,
p^{\sigma }\,
p^{\,\rho -1_{\hbox{\normalsize\it j}}}\,
\dfrac{\partial \alpha _{\sigma }}{\partial q_{j}}
\right) 
dqdp
= 
\\
& =
-
\sum\limits_{\sigma ,\rho }\sum\limits_{j}
\int
A_{\sigma +\rho -1_{\hbox{\normalsize\it j}}}\,
\left( 
\sigma _{j}
\alpha_{\sigma }
\dfrac{\partial \beta _{\rho }}{\partial q_{j}}
-
\rho _{j}\beta_{\rho }
\dfrac{\partial \alpha _{\sigma }}{\partial q_{j}}
\right) 
dq
=
\\
& =:
\sum\limits_{\sigma ,\rho }\sum\limits_{j}
\left\langle
A_{\sigma +\rho -1_{\hbox{\normalsize\it j}}}\,,
\left( 
\mathrm{ad}_{\beta_{\rho }}
\right) 
_{j}
\alpha _{\sigma }
\right\rangle 
= 
\\
& =
-
\sum\limits_{\sigma ,\rho }\sum\limits_{j}
\int 
\left[
\rho _{j}
\beta _{\rho }
\dfrac{\partial }{\partial q_{j}}
A_{\sigma +\rho -1_{\hbox{\normalsize\it j}}}\,
+\left( 
\sigma _{j}+\rho _{j}\right) 
A_{\sigma+\rho -1_{\hbox{\normalsize\it j}}}\,
\dfrac{\partial \beta _{\rho }}{\partial q_{j}}
\right] 
\alpha _{\sigma }
dq
= 
\\
& =:
-
\sum\limits_{\sigma ,\rho }\sum\limits_{j}
\left\langle 
\left( 
\mathrm{ad}_{\beta _{\rho }}^{\ast }
\right) _{j}
A_{\sigma +\rho -1_{\hbox{\normalsize\it j}}\,},
\alpha _{\sigma}
\right\rangle 
\end{align*}
where the sum is extended to all $\sigma ,\rho \in \mathbb{N}^{3}$ and
we have introduced the notation,
\begin{equation*}
_{1_{\hbox{\normalsize\it j}}\,
=:
\,
(0,...\underset{\underset{\overbrace{
j^{th}\,\text{element}}}{\uparrow }}{,1,}...,0)}
\end{equation*}
so that $\left( 1_{j}\right) _{i}=\delta _{ji}$.

The LPV bracket in terms of the $p$-moments\ may then be written as 
\begin{equation*}
\frac{\partial A_{\sigma }}{\partial t}
=
-
\sum\limits_{\rho \in \mathbb{N}^{3}}
\sum\limits_{j}
\left( 
\mathrm{ad}_{\frac{\delta h}{\delta A_{\rho }}}^{\ast }
\right) 
_{j}
A_{\sigma +\rho +1_{\hbox{\normalsize\it j}}}
\end{equation*}
where the Lie bracket is now expressed as 
\begin{equation*}
\left[\!\! 
\left[ 
\frac{\delta g}{\delta A_{\sigma }},
\frac{\delta h}{\delta A_{\rho }}
\right]\!\! 
\right] 
_{\hbox{\normalsize\it j}}
=
\sigma _{j}
\alpha _{\sigma }
\dfrac{\partial }{\partial q_{j}}
\frac{\delta h}{\delta A_{\rho }}
-
\rho _{j}
\beta _{\rho }
\dfrac{\partial }{\partial q_{j}}
\frac{\delta g}{\delta A_{\sigma }}.
\end{equation*}
Moreover the evolution of a particular $p$-moment $A_{\sigma }$ is obtained
by 
\begin{align*}
\frac{\partial A_{\sigma }}{\partial t}
& =
\left\{ 
A_{\sigma },h
\right\} 
= 
\\
& =
-
\sum\limits_{\rho}
\sum\limits_{j}
\left[ 
\rho _{j}
\frac{\delta h}{\delta A_{\rho }}
\dfrac{\partial }{\partial q_{j}}
A_{\sigma+\rho -1_{\hbox{\normalsize\it j}}}
+
\left( 
\sigma _{j}+\rho _{j}
\right)
A_{\sigma +\rho -1_{\hbox{\normalsize\it j}}}\,
\dfrac{\partial }{\partial q_{j}}
\frac{\delta h}{\delta A_{\rho }}
\right]
\end{align*}
and the KMLP bracket among the multi-dimensional $p-$moments is given in
by
\begin{equation*}
\left\{ 
A_{\sigma },A_{\rho }
\right\} 
=
-
\sum\limits_{j}
\left( 
\sigma _{j}
\frac{\partial }{\partial q_{j}}
A_{\sigma +\rho -1_{\hbox{\normalsize\it j}}}\,
+
\rho _{j}
A_{\sigma +\rho -1_{\hbox{\normalsize\it j}}}\,
\frac{\partial }{\partial q_{j}}
\right).
\end{equation*}
Inserting the previous operator in this multi-dimensional KMLP bracket
leads to
\begin{equation*}
\left\{ 
g,h
\right\} 
\left( 
\left\{ 
A
\right\} 
\right) 
=
\sum\limits_{\sigma,\rho}
\int 
\frac{\delta g}{\delta A_{\sigma }}
\left\{
A_{\sigma },
A_{\rho }
\right\} 
\frac{\delta h}{\delta A_{\rho }}
dq
\end{equation*}
and the corresponding evolution equation becomes
\begin{equation*}
\frac{\partial A_{\sigma }}{\partial t}
=
\sum\limits_{\rho}
\left\{ 
A_{\sigma },
A_{\rho }
\right\} 
\frac{\delta h}{\delta A_{\rho }}.
\end{equation*}
Thus, in multi-index notation, the form of the Hamiltonian evolution
under the KMLP bracket is essentially unchanged in going to higher
dimensions. 

\subsection{Applications of the KMLP bracket} 

The KMLP bracket was derived in the context of Benney long waves,
whose Hamiltonian is
\begin{equation*}
H_2=\frac12 (A_2+A_0^2).
\end{equation*}
This leads to the moment equations
\begin{equation*}
\frac{\partial A_n}{\partial t}
+
\frac{\partial A_{n+1}}{\partial q}
+
n A_{n-1}
\frac{\partial A_0}{\partial q}
=0
\end{equation*}
derived by Benney \cite{Be1973} as a description of long waves on a
shallow perfect fluid, with a free surface at $y=h(q,t)$. In his
interpretation, the $A_n$ were vertical moments of the horizontal
component of the velocity $p(q,y,t)$:
\begin{equation*}
A_n=\int_{y=0}^{h} p(q,y,t)^n \,\text{d}y.
\end{equation*}
The corresponding system of evolution equations for $p(q,y,t)$ and $h(q,t)$
is related by hodograph transformation, $y=\int_{-\infty}^p f(q,p',t)\,
\text{d}p'$, to the Vlasov equation
\begin{equation*}
\frac{\partial f}{\partial t}
+
p
\frac{\partial f}{\partial q}
-
\frac{\partial A_0}{\partial q}
\frac{\partial f}{\partial p}
=0.
\end{equation*}
The most important fact about the Benney hierarchy is that it is completely
integrable. This fact emerges from the following observation. Upon defining a
function $\lambda(q,p,t)$ by the principal value integral, 
\begin{equation*}
\lambda(q,p,t)=
p+P\int_{-\infty}^\infty \frac{f(q,p\,',t)}{p-p\,'} \,\text{d}p\,',
\end{equation*}
it is straightforward to verify \cite{LeMa} that
\begin{equation*}
\frac{\partial \lambda}{\partial t}
+
p
\frac{\partial \lambda}{\partial q}
-
\frac{\partial A_0}{\partial q}
\frac{\partial \lambda}{\partial p}
=0;
\end{equation*}
so that $f$ and $\lambda$ are advected along the same characteristics.

In higher dimensions, particularly $n=3$, we may take the direct sum of the KMLP bracket, together with with the Poisson bracket for an electromagnetic field (in the Coulomb gauge) where the electric field $\mathbf{E}$ and magnetic vector potential $\mathbf{A}$ are canonically conjugate; then the Hamiltonian

\begin{equation*} 
H_{MV} =\int \int \left[\frac{1}{2m} |\mathbf{p} - e\mathbf{A}|^2 \right] f(\mathbf{p},\mathbf{q}) \mathrm{d}^n \mathbf{p} \mathrm{d}^n \mathbf{q} + \int \left[\frac{1}{2} |\mathbf{E}|^2 +  \frac{1}{4}\sum_{i=1}^n
\sum_{j=1}^n (A_{i,j}-A_{j,i})^2 \right]   \mathrm{d}^n \mathbf{q} 
\end{equation*} 
yields the Maxwell-Vlasov (MV) equations for systems of interacting
charged particles. For a discussion of the MV equations from a geometric
viewpoint in the same spirit as the present approach, see \cite{CeHoHoMa1998}. For 
discussions of the Lie-algebraic approach to the control and  steering of
charged particle beams, see 
\cite{Dragt}. 

\section{Variational principles and Hamilton-Poincar\'e formulation}
\label{VarPrincHP}
In this section we show how the $p-$moment dynamics can be derived from
Hamilton's principle both in the Hamilton-Poincar\'e and Euler-Poincar\'e
forms. These variational principles are defined , respectively, on the
dual Lie algebra $\mathfrak{g}^{\ast}$ containing the moments, and on the
Lie algebra $\mathfrak{g}$ itself. For further details about these dual
variational formulations, see \cite{CeMaPeRa} and \cite{HoMaRa}. Summation
over repeated indices is intended in this section

\subsection{Hamilton-Poincar\'e hierarchy} We begin with the Hamilton-Poincar\'{e} principle for the $p-$moments written
as%
\[
\delta
\int_{t_{i}}^{t_{j}}
dt
\left(  
\left\langle 
A_{n},
\beta_{n}
\right\rangle 
-
H
\left(  
\left\{
A
\right\}  
\right)  
\right)  
=
0
\]
(where $\beta_{n}\in\mathfrak{g}$). We shall prove that this leads
to the same dynamics as found in the context of the KMLP bracket. To this
purpose, we must define the $n-$th $p-$moment in terms of the Vlasov
distribution function. We check that%
\begin{align*}
0 &  =\delta
\int_{t_{i}}^{t_{j}}
dt
\left(  
\left\langle 
A_{n},
\beta_{n}
\right\rangle 
-
H
\left(  
\left\{
A
\right\}  
\right)  
\right)  
=
\\
&  =
\int_{t_{i}}^{t_{j}}
dt
\left(  
\delta
\left\langle\! 
\left\langle 
f,
p^{n}
\beta_{n}
\right\rangle\!
\right\rangle 
-
\left\langle\!\!\! 
\left\langle 
\delta f,
\dfrac{\delta H}{\delta f}
\right\rangle\!\!\! 
\right\rangle 
\right)  
=
\\
&  =
\int_{t_{i}}^{t_{j}}
dt
\left(  
\left\langle\!\!\! 
\left\langle 
\delta f,
\left(  
p^{n}
\beta_{n}
-
\dfrac{\delta H}{\delta f}
\right)  
\right\rangle\!\!\! 
\right\rangle 
+
\left\langle\!
\left\langle 
f,
\delta
\left(  
p^{n}
\beta_{n}
\right)  
\right\rangle\!
\right\rangle 
\right)
\end{align*}

Now recall that any $g=\delta G/\delta f$ belonging to the Lie algebra
$\mathfrak{s}$ of the symplectomorphisms (which also contains the
distribution function itself) may be expressed as
\begin{align*}
g  &  =
\frac{\delta G}{\delta f}=
p^{m}
\frac{\delta G}{\delta A_{m}}=
p^{m}\xi_{m}
\end{align*}
by the chain rule. Consequently, one finds the pairing relationship,
\[
\left\langle\!\!\! 
\left\langle 
\delta f,
\left(  
p^{n}
\beta_{n}
-
\dfrac{\delta H}{\delta f}
\right)  
\right\rangle\!\!\! 
\right\rangle 
=
\left\langle 
\delta A_{n},
\left(  
\beta_{n}
-
\dfrac{\delta H}{\delta A_{n}}
\right)  
\right\rangle
\]
Next, recall from the general theory that variations on a Lie group
induce variations on its Lie algebra of the form
\[
\delta w
=
\dot{u}
+
\left[  
g,u
\right]
\]
where $u,w\in\mathfrak{s}$ and $u$ vanishes at the endpoints. Writing
$u=p^{m}\eta_{m}$ then leads to
\begin{align*}
\int_{t_{i}}^{t_{j}}
dt
\left\langle\! 
\left\langle 
f,
\delta
\left(  
p^{n}\beta_{n}
\right)
\right\rangle\! 
\right\rangle  
&  =
\int_{t_{i}}^{t_{j}}
dt
\left\langle\! 
\left\langle 
f,
\left(  
\dot{u}
+
\left[
p^{n}
\beta_{n},
u
\right]  
\right)  
\right\rangle\! 
\right\rangle 
=\\
&  =
-
\int_{t_{i}}^{t_{j}}
dt
\left(  
\left\langle 
\dot{A}_{m},
\eta_{m}
\right\rangle 
-
\left\langle
A_{n+m-1},
\left[\!
\left[  
\beta_{n},
\eta_{m}
\right]\!
\right]  
\right\rangle
\right)  
=\\
&  =-
\int_{t_{i}}^{t_{j}}
dt
\left\langle 
\left(  
\dot{A}_{m}
+\mathrm{ad}_{\beta_{n}}^{\ast}
A_{m+n-1}
\right),
\eta_{m}
\right\rangle
\end{align*}
Consequently, the Hamilton-Poincar\'{e} principle may be written entirely
in terms of the moments as
\[
\delta S=
\int_{t_{i}}^{t_{j}}
dt
\left\{  
\left\langle 
\delta A_{n},
\left(  
\beta_{n}
-
\dfrac{\delta H}{\delta A_{n}}
\right)  
\right\rangle 
-\left\langle 
\left(  
\dot{A}_{m}
+
\mathrm{ad}_{\beta_{n}}^{\ast}
A_{m+n-1}
\right),
\eta_{m}
\right\rangle 
\right\}  
=
0
\]
This expression produces the inverse Legendre transform%
\[
\beta_{n}=\dfrac{\delta H}{\delta A_{n}}%
\]
(holding for hyperregular Hamiltonians). It also yields the equations of
motion
\[
\frac{\partial A_{m}}{\partial t}=-\mathrm{ad}_{\beta_{n}}^{\ast}A_{m+n-1}
\]
which are valid for arbitrary variations $\delta A_{m}$ and variations
$\delta\beta_{m}$ of the form
\[
\delta\beta_{m}=\dot{\eta}_{m}+\mathrm{ad}_{\beta_{n}}\eta_{m-n+1}
\]
where the variations $\eta_{m}$ satisfy vanishing endpoint conditions,
\[
\left.  \eta_{m}\right\vert _{t=t_{i}}=\left.  
\eta_{m}\right\vert _{t=t_{j}}=0
\]
Thus, the Hamilton-Poincar\'e variational principle recovers the
hierarchy of the evolution equations derived in the previous section
using the KMLP bracket.

\subsection{Euler-Poincar\'e hierarchy} 
The corresponding Lagrangian formulation of the Hamilton's principle now
yields
\begin{align*}
\delta
\int_{t_{i}}^{t_{j}}
L
\left(  
\left\{  
\beta
\right\}  
\right)  
dt 
&  =
\int_{t_{i}}^{t_{j}}
\left\langle 
\frac{\delta L}{\delta\beta_{m}},
\delta\beta_{m}
\right\rangle
dt
=\\
&  =
\int_{t_{i}}^{t_{j}}
\left\langle 
\frac{\delta L}{\delta\beta_{m}},
\left(  
\dot{\eta}_{m}
+\mathrm{ad}_{\beta_{n}}
\eta_{m-n+1}
\right)  
\right\rangle 
dt
=\\
&  =
-
\int_{t_{i}}^{t_{j}}
\left(  
\left\langle 
\frac{\partial}{\partial t}
\frac{\delta L}{\delta \beta_{m}},
\eta_{m}
\right\rangle 
+\left\langle 
\mathrm{ad}_{\beta_{n}}^{\ast}
\frac{\delta L}{\delta\beta_{m}}
,\eta_{m-n+1}
\right\rangle 
\right)
dt
=\\
&  =
-
\int_{t_{i}}^{t_{j}}
\left(  
\left\langle 
\frac{\partial}{\partial t}
\frac{\delta L}{\delta \beta_{m}},
\eta_{m}
\right\rangle 
+
\left\langle 
\mathrm{ad}_{\beta_{n}}^{\ast}
\frac{\delta L}{\delta\beta_{m+n-1}},
\eta_{m}
\right\rangle 
\right)
dt
=\\
&  =
-
\int_{t_{i}}^{t_{j}}
\left\langle 
\left(  
\frac{\partial}{\partial t}
\frac{\delta L}{\delta
\beta_{m}}
+
\mathrm{ad}_{\beta_{n}}^{\ast}
\frac{\delta L}{\delta \beta_{m+n-1}}
\right),
\eta_{m}
\right\rangle 
dt
\end{align*}
upon using the expression previously found for the variations
$\delta\beta_{m}$ and relabeling indices appropriately.
The Euler-Poincar\'e equations may then be written as

\[
\frac{\partial}{\partial t}
\frac{\delta L}{\delta \beta_{m}}
+
\mathrm{ad}_{\beta_{n}}^{\ast}
\frac{\delta L}{\delta \beta_{m+n-1}}
=
0
\]
with the same constraints on the variations as in the previous paragraph.
Applying the Legendre transformation
\[
A_{m}=\dfrac{\delta L}{\delta\alpha_{m}}%
\]
yields the Euler-Poincar\'e equations (for hyperregular Lagrangians). This
again leads to the same hierarchy of equations derived earlier using the
KMLP bracket.

To summarize, the calculations in this section have proven the following result.
\begin{theorem}
With the above notation and hypotheses of hyperregularity the following
statements are equivalent:

\begin{enumerate}
\item The Euler--Poincar\'{e} Variational Principle. The curves $\beta_{n}(t)$ are
critical points of the action
\[
\delta
\int_{t_{i}}^{t_{j}}
L\left(  \left\{  \beta\right\}  \right)  dt=0
\]
for variations of the form
\[
\delta\beta_{m}=\dot{\eta}_{m}+\mathrm{ad}_{\beta_{n}}\eta_{m-n+1}
\]
in which $\eta_{m}$ vanishes at the endpoints
\[
\left.  \eta_{m}\right\vert _{t=t_{i}}=\left.  \eta_{m}\right\vert _{t=t_{j}}=0
\]
and the variations $\delta A_{n}$ are arbitrary.

\item The Lie--Poisson Variational Principle. The curves $(\beta_{n}
,A_{n})\left(  t\right)  $ are critical points of the action
\[
\delta
\int_{t_{i}}^{t_{j}}
\left(  
\left
\langle A_{n},
\beta_{n}
\right\rangle 
-
H
\left(  
\left\{
A
\right\}  
\right)  
\right)  
dt
=
0
\]
for variations of the form
\[
\delta\beta_{m}=\dot{\eta}_{m}+\mathrm{ad}_{\beta_{n}}\eta_{m-n+1}
\]
where $\eta_{m}$ satisfies endpoint conditions
\[
\left.  \eta_{m}\right\vert _{t=t_{i}}=\left.  \eta_{m}\right\vert _{t=t_{j}}=0
\]
and where the variations $\delta A_{n}$ are arbitrary.

\item The Euler--Poincar\'{e} equations hold:
\[
\frac{\partial}{\partial t}
\dfrac{\delta L}{\delta\beta_{m}}
+
\mathrm{ad}_{\beta_{n}}^{\ast}
\dfrac{\delta L}{\delta\beta_{m+n-1}}
=
0.
\]

\item The Lie--Poisson equations hold:%
\[
\dot{A}_{m}
=
-
\mathrm{ad}_{\frac{\delta H}{\delta A_{n}}}^{\ast}A_{m+n-1}
\]

\end{enumerate}
\end{theorem}

For further details on the proof of this theorem we address the reader to
\cite{CeMaPeRa}. An analogous result is also valid in the multidimensional
case with slight modifications.

\section{Quadratic Hamiltonians}\label{geo-prob}
\subsection{Geodesic motion}
We shall consider the problem of geodesic motion on the space of
$p-$moments. For this, we define the Hamiltonian as the norm on the
$p-$moment given by the following metric and inner product,
\begin{eqnarray}
h=\frac{1}{2}\|A\|^2
&=&
\frac{1}{2}\sum_{n,s=0}^\infty
\int\hspace{-3mm}\int
A_n(q)G_{ns}(q,q\,')A_s(q\,')\,dq\,dq\,'
\label{Ham-metric}
\end{eqnarray}
The metric $G_{ns}(q,q\,')$ is chosen to be positive definite,
so it defines a norm for $\{A\}\in\mathfrak{g}^*$. The
corresponding geodesic equation with respect to this norm is
found as in the previous section to be,
\begin{eqnarray}
\frac{\partial  A_m}{\partial t}
=
\{\,A_m\,,\,h\,\}
=
-\sum_{n=0}^\infty
\Big(n\beta_n
\frac{\partial}{\partial q} A_{m+n-1}
+
(m+n)A_{m+n-1}\frac{\partial}{\partial q}
\beta_n
\Big)
\label{EPMS-eqn}
\end{eqnarray}
with dual variables $\beta_n\in\mathfrak{g}$ defined by
\begin{eqnarray}
\beta_n
=
\frac{\delta h}{\delta A_n} 
=
\sum_{s=0}^\infty
\int
G_{ns}(q,q\,')A_s(q\,')\,dq\,'
=
\sum_{s=0}^\infty
G_{ns}*A_s
\,.
\label{EPMS-vel}
\end{eqnarray}
Thus, evolution under (\ref{EPMS-eqn}) may be rewritten as
coadjoint motion on $\mathfrak{g}^*$
\begin{eqnarray}
\frac{\partial  A_m}{\partial t}
=
\{\,A_m\,,\,h\,\}
=:
-\sum_{n=0}^\infty
{\rm ad}^*_{\beta_n}A_{m+n-1}
\label{A-dot}
\end{eqnarray}
This system comprises an infinite system of nonlinear, nonlocal, coupled
evolutionary equations for the $p-$moments. In this system, evolution of
the $m^{th}$ moment is governed by the potentially infinite sum of
contributions of the velocities $\beta_n$ associated with $n^{th}$
moment sweeping the $(m+n-1)^{th}$ moment by coadjoint action.
Moreover, by equation (\ref{EPMS-vel}), each of the $\beta_n$
potentially depends nonlocally on all of the moments. 

Equations (\ref{Ham-metric}) and (\ref{EPMS-vel}) may be
written in three dimensions in multi-index notation, as follows:
the Hamiltonian is given by
\[
h
=
\frac{1}{2}
\left\vert 
\left\vert 
A
\right\vert 
\right\vert 
^{2}
=
\frac{1}{2}
\sum\limits_{\mu,\nu}
\iint
A_{\mu}
\left(  
\mathbf{q},
t
\right)  
G_{\mu\nu}
\left(  
\mathbf{q,q}\,^{\prime}
\right)  
A_{\nu}
\left(  
\mathbf{q}\,^{\prime},
t
\right)  
d
\mathbf{q}
d
\mathbf{q}\,^{\prime}
\]
so the dual variable is written as
\begin{align*}
\beta_{\rho}
=
\frac{\delta h}{\delta A_{\rho}}  
& =
\sum\limits_{\nu}
\iint
G_{\rho\nu}
\left(  
\mathbf{q,q}\,^{\prime}
\right)  
A_{\nu}
\left(  
\mathbf{q}\,^{\prime},
t
\right)  
d\mathbf{q}
d\mathbf{q}\,^{\prime}
=
\sum\limits_{\nu}
G_{\rho\nu}\ast A_{\nu}.
\end{align*}
\\

\subsection{Singular geodesic solutions}
Remarkably, in any number of spatial dimensions, the geodesic
equation (\ref{EPMS-eqn}) possesses exact solutions which are {\it
singular}; that is, they are supported on delta functions in
$q-$space. 

\begin{theorem}[Singular solution Ansatz for geodesic flows of Vlasov
$p-$moments]$\quad$\\   Equation (\ref{EPMS-eqn}) admits singular
solutions of the form
\begin{eqnarray}
A_\sigma(\mathbf{q},t)
&=&
\sum_{j=1}^N
\int
P_j^\sigma(\mathbf{q},t,a_j)\,
\delta\big(\mathbf{q}-\mathbf{Q}_j(t,a_j)\big)\,da_j
\label{sing-soln}
\end{eqnarray}
in which the integrals over coordinates $a_j$ are performed over
$N$ embedded subspaces of the $q-$space and the parameters
$(Q_j,P_j)$ satisfy canonical Hamiltonian equations in
which the Hamiltonian is the norm $h$ in (\ref{Ham-metric})
evaluated on the singular solution Ansatz (\ref{sing-soln}). 
\end{theorem}

In one dimension, the coordinates $a_j$ are absent and the
singular solutions in (\ref{sing-soln}) reduce to
\begin{eqnarray}
A_s(q,t)
&=&
\sum_{j=1}^N
P_j^s(q,t)\,
\delta\big(q-Q_j(t)\big).
\label{sing-soln-1D}
\end{eqnarray}
In order to show this is a solution in one dimension, one checks that
these singular solutions satisfy a system of partial differential
equations in Hamiltonian form, 
whose Hamiltonian couples all the moments
\begin{eqnarray*}
H_N
&=&
\frac{1}{2}\sum_{n,s=0}^\infty
\sum_{j,k=1}^N
P^s_j(Q_j(t),t)P^n_k(Q_k(t),t)\,
G_{ns}(Q_j(t),Q_k(t))
\end{eqnarray*}
One forms the pairing of the coadjoint equation
\[
\dot{A}_{m}
=
-\sum_{n,s}\mathrm{ad}_{G_{ns}\ast A_{s}}^{\ast}A_{m+n-1}
\]
with a sequence of smooth functions $\left\{  \varphi_{m}\left(  q\right) 
\right\} 
$, so that:
\[
\langle \dot{A}_{m},\varphi_{m}\rangle 
=
\sum_{n,s}
\left\langle A_{m+n-1},\mathrm{ad}_{G_{ns}\ast A_{s}}\varphi_{m}\right\rangle
\]
One expands each term and denotes $\widetilde{P}_{j}(t):=P_{j}(Q_j,t)$:
\begin{align*}
\langle \dot{A}_{m},\varphi_{m}\rangle  
&  =
\sum_{j}\int dq\,
\varphi_{m}\left(q\right) \frac{\partial}{\partial t}\left[  P_{j}^{m}\left(q,t\right)  \delta\left(  q-Q_{j}  \right)  \right]  
=\\
&  =
\sum_{j}\int dq
\varphi_{m}\left(q\right)  
\left[
\delta\left(q-Q_j\right)  
\frac{\partial P_{j}^{m}}{\partial t}-P_{j}^{m}\dot Q_{j}
\delta\,^{\prime}\left(  q-Q_{j}\right)  
\right]  
=\\
&  =
\sum_{j}
\left(  
\frac{d\widetilde{P}_{j}^{m}}{d t}\varphi_{m}\left(Q_{j}\right)  
+
\widetilde{P}_{j}^{m}\dot Q_{j}
\varphi_{m}\,^{\prime}\left(  Q_{j}\right)  
\right)
\end{align*}
Similarly expanding
\begin{align*}
\left\langle A_{m+n-1},\mathrm{ad}_{G_{ns}\ast A_{s}}\varphi_{m}\right\rangle
&  =
\sum_{j,k}\int dq\,
\widetilde{P}_{k}^{s}\, P_j^{m+n-1}\delta\left(  q-Q_{j}\right) \left[  
n\varphi_{m}\,^{\prime}G_{ns}\left( q,Q_{k}\right)  
-
m\varphi_{m}\frac{\partial G_{ns}\left(q,Q_{k}\right)}{\partial q}
\right]  
=\\
&  =
\sum_{j,k}\widetilde{P}_{k}^{s}\widetilde{P}_{j}^{m+n-1}
\left[
n\,\varphi_{m}\,^{\prime}\left(  Q_{j}\right)G_{ns}\left(Q_{j},Q_{k}\right)
-
m\,\varphi_{m}\left(Q_{j}\right)
\frac{\partial G_{ns}\left(Q_{j},Q_{k}\right)}{\partial Q_{j}}
\right]
\end{align*}
leads to
\begin{align*}
\widetilde{P}_{j}^{m}\frac{d Q_{j}}{d t} 
&  =
\sum_{n,s}\sum_{k}
n\,\widetilde{P}_{k}^{s}\,\widetilde{P}_{j}^{m+n-1}G_{ns}\left(Q_{j},Q_{k}\right)
\\
\frac{d\widetilde{P}_{j}^{m}}{d t} 
&  =
-
m\sum_{n,s}\sum_{k}\widetilde{P}_{k}^{s}\,\widetilde{P}_{j}^{m+n-1}
\frac{\partial G_{ns}\left(Q_{j},Q_{k}\right)}{\partial Q_{j}}
\end{align*}
so that we finally obtain equations for $Q_{j}$ and $\widetilde{P}_{j}$ in
canonical form, 
\[
\frac{d Q_{j}}{d t}
=
\frac{\partial H_N}{\partial\widetilde{P}_{j}},
\qquad
\frac{d\widetilde{P}_{j}}{d t} 
=
-\,\frac{\partial H_N}{\partial Q_{j}}.
\]

\paragraph{Remark about higher dimensions}
The singular solutions (\ref{sing-soln}) with the
integrals over coordinates $a_j$ exist in higher
dimensions. The higher dimensional singular solutions satisfy a system of
canonical Hamiltonian integral-partial differential equations, instead of
ordinary differential equations. 

\section{Discussion}\label{geo-singsoln}

\subsection{Remarks about EPSymp and connections with EPDiff}
Importantly, geodesic motion for the $p-$moments is equivalent to
geodesic motion for the Euler-Poincar\'e equations on the
symplectomorphisms (EPSymp) given by the following Hamiltonian

\begin{equation}\label{epsymp}
H\left[  f\right]  =
\frac{1}{2}\iint f\left(q,p,t\right)  
\mathcal{G}\left(  q,p,q\,^{\prime},p\,^{\prime}\right)  
f\left(q\,^{\prime},p\,^{\prime},t\right)  
dq\,dp\,dq\,^{\prime}dp\,^{\prime}
\end{equation}
The equivalence with EPSymp emerges when the function $\mathcal{G}$
is written as

\[
\mathcal{G}\left(  q,q\,^{\prime},p,p\,^{\prime}\right)  
=
\underset{n,m}{\sum}
\thinspace 
p^{n}\/G_{nm}\left(q,q\,^{\prime}\right) p\,^{\prime\, m}
\,.
\]
Thus, whenever the metric $\mathcal{G}$ for EPSymp has a Taylor series,
its solutions may be expressed in terms of the geodesic motion for the
$p-$moments.

Moreover the distribution function corresponding to the singular solutions
for the moments is a particular case of the {\bfi cold-plasma
approximation}, given by

\[
f(q,p,t)=\sum_j \rho_j(q,t)\,\delta(p-P_j(q,t))
\]
where in our case a summation is introduced and  $\rho$ is written as a Lagrangian particle-like
density:

\[
\rho_j(q,t)=\delta(q-Q_j(t))
\]
To check this is a solution for the geodesic motion of the generating function, one repeats exactly the same procedure as for the moments, in order to find the following Hamiltonian equations

\[
\frac {d Q_j}{d t}=\frac \partial {\partial \widetilde{P}_j}\,\,
\frac {\delta H}{\delta f} ( Q_j,\widetilde{P}_j ),
\qquad
\frac {d \widetilde{P}_j}{d t}=\frac \partial {\partial Q_j}\,\,
\frac {\delta H}{\delta f} ( Q_j,\widetilde{P}_j )
\]
where $\widetilde{P}_j=P_j\circ Q_j$ denotes the composition of the two functions
$P_j$ and $Q_j$. This procedure recovers single particle motion for density $\rho_j$ defined on a delta function.
\\
As we shall show below, these singular solutions of EPSymp are also
solutions of the Euler-Poincar\'e equations on the diffeomorphisms
(EPDiff), provided one truncates to consider only first order moments
\cite{HoMa2004}. With this truncation, the singular solutions in the case
of single-particle dynamics reduce in one dimension to the pulson
solutions for EPDiff
\cite{CaHo1993}. 

\subsection{Exchanging variables in EPSymp}
One can show that exchanging the variables $q\leftrightarrow p$ in the
single particle PDF leads to another nontrivial singular
solution of EPSymp, which is different from those found previously. To
see this, let
$f$ be given by

\[
f(q,p,t)=\sum_j\delta(q-Q_j(p,t))\,\delta(p-P_j(t))
\]
At this stage nothing has changed with respect to the previous solution since
the generating function is symmetric with respect to q and p. However,
inserting this expression in the definition of the $m-$th moment yields

\[
A_m(q,t)=\sum_j\, P_j^m\,\delta(q-Q_j(P_j,t))
\] 
which is quite different from the solutions found previously. One
again obtains a canonical Hamiltonian structure for $P_j$ and $Q_j$.\\
This second expression is an alternative parametrisation of the cold-plasma
reduction above and it may be useful in situations where the composition
$Q_j\circ P_j$ is more convenient than $P_j\circ Q_j$.

\subsection{Remarks about truncations}
The problem presented by the coadjoint motion equation  (\ref{A-dot}) for
geodesic evolution of $p-$moments under EPDiff needs further
simplification. One simplification would be to truncate the (doubly)
infinite set of equations in  (\ref{A-dot}) to a finite set. These moment
dynamics may be truncated at any stage. For example, two alternatives could
be
\begin{itemize}
\item
by modifying the Hamiltonian, so that some moments
are decoupled (or, perhaps so that some moments slaved to
others). For example, one might also arrange for the metric
$G_{ns}$ 
\begin{itemize}
\item
to only couple nearest neighbors, or 
\item
to be diagonal, or
\item
to be a multiple of the identity
\end{itemize}
\item
by modifying the Lie-algebra in the KMLP bracket to
vanish for weights $m+n-1$ greater than a chosen cut-off value.
\end{itemize}

\subsection{Examples of simplifying truncations and specializations.} 
For example, if we truncate the sums to $m,n=0,1,2$ only, then
equation (\ref{A-dot}) produces the coupled system of partial
differential equations,
\begin{align*}
\frac{\partial A_{0}}{\partial t} &  =-\mathrm{ad}_{\beta_{1}}^{\ast}
A_{0}-\mathrm{ad}_{\beta_{2}}^{\ast}A_{1}\,,\\
\frac{\partial A_{1}}{\partial t} &  =-\mathrm{ad}_{\beta_{0}}^{\ast}
A_{0}-\mathrm{ad}_{\beta_{1}}^{\ast}A_{1}-\mathrm{ad}_{\beta_{2}}^{\ast}
A_{2}\,,\\
\frac{\partial A_{2}}{\partial t} &  =-\mathrm{ad}_{\beta_{0}}^{\ast}
A_{1}-\mathrm{ad}_{\beta_{1}}^{\ast}A_{2}\,.
\end{align*}
Expanding now the expression of the coadjoint operation
\[
\mathrm{ad}_{\beta_{h}}^{\ast}A_{k+h-1}=\left(  k+h\right)  A_{k+h-1}
\partial_{q}\beta_{h}+h\beta_{h}\partial_{q}A_{k+h-1}
\]
and relabeling
\[
\mathrm{ad}_{\beta_{h}}^{\ast}A_{k}=\left(  k+1\right)  A_{k}\partial_{q}
\beta_{h}+h\beta_{h}\partial_{q}A_{k}
\]
one calculates
\begin{align*}
\frac{\partial A_{0}}{\partial t} &  =-\partial_{q}\left(  A_{0}\beta_{1}\right)  -2A_{1}\partial_{q}\beta
_{2}-2\beta_{2}\partial_{q}A_{1}\\
\frac{\partial A_{1}}{\partial t} &  =-A_{0}\partial_{q}\beta_{0}-2A_{1}\partial_{q}\beta_{1}-\beta_{1}%
\partial_{q}A_{1}-3A_{2}\partial_{q}\beta_{2}-2\beta_{2}\partial_{q}A_{2}\\
\frac{\partial A_{2}}{\partial t} &  =-2A_{1}\partial_{q}\beta_{0}-3A_{2}\partial_{q}\beta_{1}-\beta_{1}%
\partial_{q}A_{2}%
\end{align*}

We specialize to the case that each velocity depends only on its
corresponding moment, so that $\beta_s=G*A_s$, $s=0,1,2$. If we further
specialize by setting $A_0$ and $A_2$ initially to zero, then these three
equations reduce to the single equation 
\begin{eqnarray*}
\frac{\partial  A_1}{\partial t}
&=&
-\,\beta_1\,{\partial_q}A_1
-\,2A_1\,{\partial_q}\beta_1
\,.
\end{eqnarray*}
Finally, if we assume that $G$ in the convolution $\beta_1=G*A_1$ is the
Green's function for the operator relation 
\[
A_1=(1-\alpha^2\partial_q^2)\beta_1
\]
for a constant lengthscale $\alpha$, then the evolution equation
for $A_1$ reduces to the integrable Camassa-Holm (CH) equation
\cite{CaHo1993} in the absence of linear dispersion. This is the one-dimensional
EPDiff equation, which has singular (peakon) solutions. Thus, after these
various specializations of the EPDiff $p-$moment equations, one finds the
integrable CH peakon equation as a specialization of the coadjoint moment
dynamics of equation (\ref{A-dot}).

That such a strong restriction of the $p-$moment system leads to
such an interesting special case bodes well for future investigations of
the EPSymp $p-$moment equations. Further specializations and truncations
of these equations will be explored elsewhere. Before closing, we mention
one or two other open questions about the solution behavior of the
$p-$moments of EPSymp.

\subsection{Open questions for future work}
Several open questions remain for future work. The first of these is whether the singular solutions found here will emerge
spontaneously in EPSymp dynamics from a smooth initial Vlasov PDF. This
spontaneous emergence of the  singular solutions does occur for EPDiff.
Namely, one sees the singular solutions of EPDiff emerging from {\it any}
confined initial distribution of the dual variable. (The dual variable is
fluid velocity in the case of EPDiff). In fact, integrability of EPDiff in
one dimension by the inverse scattering transform shows that {\it only}
the singular solutions (peakons) are allowed to emerge from any confined
initial distribution in that case \cite{CaHo1993}. In higher dimensions, numerical
simulations of EPDiff show that again only the singular solutions emerge
from confined initial distributions. In contrast, the point vortex
solutions of Euler's fluid equations (which are isomorphic to the cold
plasma singular solutions of the Vlasov Poisson equation) while
comprising an invariant manifold of singular solutions, do not
spontaneously emerge from smooth initial conditions in Euler fluid
dynamics. Nonetheless, something quite analogous to the singular solutions
is seen experimentally for cold plasma in a Malmberg-Penning trap \cite{DuON1999}.
Therefore, one may ask which outcome will prevail for the singular
solutions of EPSymp. Will they emerge from a confined smooth initial
distribution, or will they only exist as an invariant manifold for
special initial conditions? Of course, the interactions of these singular
solutions for various metrics and the properties of their collective
dynamics is a question for future work.

Geometric questions also remain to be addressed. In geometric fluid dynamics, Arnold and Khesin \cite{ArKe98} formulate the problem of symplecto-hydrodynamics, the symplectic counterpart of ordinary ideal hydrodynamics on the special diffeomorphisms SDiff. In this regard, the work of Eliashberg and Ratiu \cite{ElRa91} showed that dynamics on the symplectic group radically differs from ordinary hydrodynamics, mainly because the diameter of Symp($M$) is infinite, whenever $M$ is a compact exact symplectic manifold with a boundary.  Of course, the presence of boundaries is important  in fluid dynamics. However, generalizing a result by Shnirelman \cite{Shn85}, Arnold and Khesin point out that the diameter of SDiff($M$) is finite for any compact simply connected Riemannian  manifold $M$ of dimension greater than two.

In the case under discussion here, the situation again differs from that envisioned by Eliashberg and Ratiu. The EPSymp Hamiltonian (\ref{epsymp}) determines geodesic motion on Symp($T^*\mathbb{R}^3$), which may be regarded as the restriction of the Diff($T^*\mathbb{R}^3$) group, so that the Liouville volume is preserved. The main difference in our case is that $M=T^*\mathbb{R}^3$ is not compact, so one of the conditions for the Eliashberg--Ratiu result does not hold. Thus, one may ask, what are the geometric properties of Symp acting on a symplectic manifold which is not compact? What remarkable differences between Symp and SDiff remain to be found in such a situation?

Yet another interesting case occurs when the particles undergoing Vlasov dynamics are confined in a certain region of position space. In this situation, again the phase space is not compact, since the momentum may be unlimited. The dynamics on a bounded spatial domain descends from that on the unbounded cotangent bundle upon taking the $p$-moments of the Hamiltonian vector field. Thus, in this topological sense \emph{$p$-moments and $q$-moments are not equivalent}. In the present work, this distinction has been ignored by assuming either homogeneous or periodic boundary conditions.

\subsection*{Acknowledgements} This work was begun in preparation for a meeting at UC Berkeley in honor of Henry McKean, to whom we are grateful for interesting and encouraging discussions over many years. We are grateful to our 
colleagues Claudio Albanese and Greg Pavliotis at Imperial College London
for their advice and interest regarding this problem. We also thank Alan
Weinstein of UC Berkeley for correspondence and discussions in this
matter. CT is also grateful to the TERA Foundation for Oncological
Hadrontherapy and in particular to the working group at CERN (Geneva,
Switzerland) for the lively interest they expressed in this study, with
special regard to the perspectives that it may have on the design and
control of plasma beams in particle accelerators. The work of DDH was partially supported by US DOE, Office of Science, Applied Mathematics program of the Mathematical, Information, and Computational Sciences Division (MICS). 

\bibliographystyle{unsrt}


\end{document}